\documentclass[aps,prl,10pt,twocolumn,superscriptaddress,nofootinbib,floatfix]{revtex4-2}
\pdftrailerid{}
\usepackage{amsmath,amssymb}
\usepackage{graphicx}
\usepackage{float}
\usepackage{xcolor}
\colorlet{newtextcolor}{black}
\colorlet{newcitecolor}{blue!60!black}

\usepackage{tikz}
\usetikzlibrary{decorations.pathmorphing,decorations.markings,arrows.meta,shapes.geometric}
\usepackage[colorlinks=true,linkcolor=blue!60!black,citecolor=blue!60!black,urlcolor=blue!60!black]{hyperref}
\newcommand{\ah}{\hat a}
\newcommand{\CE}[1]{C_{E,#1}}
\newcommand{\pinf}{p_\infty}
\newcommand{\sechead}[1]{\textit{#1}}
\newcommand{\emappendix}[1]{%
  \setcounter{equation}{0}%
  \renewcommand{\theequation}{#1\arabic{equation}}%
  \renewcommand{\theHequation}{EM#1.\arabic{equation}}%
}
\newcommand{\EM}[1]{\hyperlink{em:#1}{[EM-#1]}}
\newcommand{\emhead}[2]{%
  \par\medskip\noindent%
  \emappendix{#1}%
  \hypertarget{em:#1}{}%
  \sechead{#1. #2}---\ignorespaces
}

\begin{document}

\title{Quadratic-in-spin scattering of compact bodies at fourth post-Minkowskian order}

\author{Taushif Ahmed}
\email{taushif.ahmed@ur.de}
\affiliation{Institute for Theoretical Physics, University of Regensburg, 93040 Regensburg, Germany}
\author{Syed Mehedi Hasan}
\email{syed-mehedi.hasan@ur.de}
\affiliation{Institute for Theoretical Physics, University of Regensburg, 93040 Regensburg, Germany}

\begin{abstract}
We present the fourth-post-Minkowskian scattering angle of aligned spinning compact bodies through quadratic spin at arbitrary velocity and mass ratio. We include spin-induced quadrupoles and radiation reaction, supplemented by curvature-squared spin-tidal terms.
Three-loop worldline integrals yield
closed forms in polylogarithms and
elliptic integrals.
Infrared poles cancel exactly, fixing the tail logarithm. We obtain a new post-Newtonian coefficient and third-order angular-momentum loss for general quadrupoles. The new input shifts resummed predictions for black-hole scattering.
\end{abstract}
\maketitle

\sechead{Introduction}---Gravitational-wave observations increasingly probe the masses, spins and internal structure of compact binaries~\cite{LIGOScientific:2016aoc,LIGOScientific:2025slb}. The greater precision anticipated from future ground- and space-based detectors~\cite{Punturo:2010zz,Reitze:2019iox,LISA:2017pwj} calls for equally precise two-body dynamics~\cite{Purrer:2019jcp}. The post-Minkowskian (PM) expansion, organized in powers of Newton's constant $G$ without expanding in velocity, provides such information from the scattering angle, a gauge-independent observable that can constrain bound-orbit and effective-one-body (EOB) models~\cite{Damour:2016gwp,Damour:2017zjx,Kalin:2019rwq,Buonanno:1998gg,Buonanno:2024vkx,Buonanno:2024byg}.
Complementary field-theoretic approaches include scattering amplitudes~\cite{Cheung:2018wkq,Bern:2019nnu},
worldline effective field theory~{\cite{Goldberger:2004jt,Kalin:2020mvi,Kalin:2020fhe,Mandal:2022nty}},
and worldline quantum field theory~\cite{Mogull:2020sak}.
Numerical-relativity simulations of scattering black holes~\cite{Damour:2014afa,Hopper:2022rwo,Rettegno:2023ghr,Long:2025nmj}
provide strong-field tests of these predictions.

Spin makes this problem sensitive to the nature of the bodies. Beyond spin--orbit coupling, quadratic order contains spin--spin interactions and the spin-induced mass quadrupoles. The latter are fixed for Kerr black holes but depend on the equation of state for neutron stars~\cite{Poisson:1997ha,Laarakkers:1997hb}, providing a handle on finite-size physics~{\cite{Krishnendu:2017shb,Mandal:2023lgy,Mandal:2023hqa,Mandal:2024iug}}. Quantum-field-theory methods have advanced the spinless scattering angle through fourth PM order~\cite{Bern:2021dqo,Dlapa:2021npj,Dlapa:2021vgp,Bern:2021yeh,Dlapa:2022lmu} and beyond~\cite{Driesse:2024xad,Driesse:2024feo,Driesse:2026qiz,Bern:2025zno}. At first PM order the spin dependence is known to all orders~\cite{Vines:2017hyw}; at second PM order,
it is known through fifth order in spin~\cite{Bern:2022kto}, and for one spinning body to all
orders~\cite{Aoude:2022trd}. For spinning bodies, quadratic-spin results have reached third PM order~\cite{Bern:2020buy,Liu:2021zxr,Kosmopoulos:2021zoq,Jakobsen:2021zvh,Jakobsen:2022fcj,FebresCordero:2022jts,Akpinar:2024meg},
together with the scattering waveform~\cite{Bohnenblust:2025gir} and the first quartic-spin
terms~\cite{Akpinar:2025bkt}. The fourth-order conservative and dissipative angles are known
through linear spin~\cite{Jakobsen:2023ndj,Jakobsen:2023hig}; see also Ref.~\cite{Gonzo:2026soft}. For Kerr black holes,
the angular momentum radiated at third PM order and the corresponding linear-in-loss contribution to
the fourth-order scattering angle have also been obtained at quadratic order in spin~\cite{Zhu:ZenodoKerr}. Post-Newtonian (PN) results for the quadratic-spin dynamics at N$^3$LO~\cite{Kim:2022bwv,Mandal:2022ufb} and spinning-test-body limits~\cite{Hoogeveen:2025tew,Bautista:2024agp} supply independent constraints, but the general two-body $G^4S^2$ scattering angle has remained unknown.

Here we determine this missing contribution for spins aligned or antialigned with the orbital angular momentum. We obtain the conservative and dissipative angle in closed form at arbitrary relative velocity and mass ratio. The three-loop worldline calculation includes spin-induced quadrupoles and radiation reaction. We additionally incorporate the curvature-squared spin-tidal contributions, obtained from a one-loop two-graviton-exchange calculation.
We establish the exact cancellation of infrared poles and determine the associated tail logarithm. We derive a new quadratic-spin PN coefficient, and third-PM angular-momentum loss for general quadrupoles. Numerical relativity (NR) simulations are precise enough to resolve the effect of the new input when it is incorporated into strong-field EOB resummations.


\sechead{Setup and result}---Two bodies of masses $m_1$, $m_2$, with $M=m_1+m_2$, $\nu=m_1m_2/M^2$ and
$\delta=(m_2-m_1)/M$, approach each other with relative Lorentz factor $\sigma=u_1\cdot u_2=(1+x^2)/(2x)$,
$0<x<1$, and speed $v=\sqrt{1-\sigma^{-2}}$. In the center-of-mass frame the energy is $E=M\Gamma$ with
$\Gamma=\sqrt{1+2\nu(\sigma-1)}$ and the momentum is $\pinf=m_1m_2\sqrt{\sigma^2-1}/E$. The covariant impact
parameter $b$ is orthogonal to both incoming velocities. The spins $S_i$ are parallel or antiparallel to the
orbital angular momentum. We write $a_i=S_i/m_i$, positive when parallel, and $\ah_i=a_i/b$. The coefficient
$\CE{i}$ parametrizes the leading spin-induced quadrupole relative to its Kerr value,
with $C_{E,i}=0$ for Kerr in our worldline convention.
With $c=1$ the angle is
\begin{equation}
\theta=\sum_n \theta^{(n)}\Big(\frac{GM}{b}\Big)^n ,\qquad
\theta^{(4)}=\sum_{S\in\mathcal S}\theta^{(4)}_S\, S ,
\label{eq:theta}
\end{equation}
where the 16 structures are
\begin{align*}
\mathcal S={}&\{1,\ah_1,\ah_2,\ah_1^2,\ah_1\ah_2,\ah_2^2,
 \CE{1}\ah_1^2,\CE{2}\ah_2^2\}\\
&\cup\bigcup_{i=1}^{2}\big\{C^{(2)}_{E^2,i}\ah_i^2,
C^{(0)}_{E^2S^2,i}\ah_i^2, C^{(2)}_{B^2,i}\ah_i^2,
C^{(0)}_{B^2S^2,i}\ah_i^2\big\}.
\end{align*}
The first eight terms, $\mathcal S_0$, arise from the curvature-linear worldline action; the remaining eight, $\mathcal S_{R^2}$, represent four independent curvature-squared couplings per body. 
{These spin-tidal coefficients, denoted collectively $C_{X,i}$, vanish for Kerr black holes in static conservative matching but remain independent for generic compact bodies~\EM{E}.}
{For black holes $a_i=Gm_i\chi_i$, where $\chi_i=S_i/(Gm_i^2)$ and $|\chi_i|\leq1$, so rapidly spinning $G^4S^2$ terms scale as nonspinning $G^6$ terms}~\cite{Jakobsen:2023ndj}.

At fourth PM order the mass dependence of each spin
structure can be organized as~\cite{Damour:2019lcq,Jakobsen:2023hig},
\begin{align}
\frac{\theta^{(4)}}{\Gamma}
={}&\sum_{S\in\mathcal S_0}S
\sum_{k,j=0}^{1}\nu^k\delta^j
\Big(c^S_{kj}+\frac{\nu}{\Gamma^2}\,r^S_{kj}\Big)
\nonumber\\
&+\nu\sum_{S\in\mathcal S_{R^2}}S
\big(c^S_{10}+\delta c^S_{11}\big).
\label{eq:mass}
\end{align}
{Each $S\in\mathcal S_0$ has four conservative functions $c^S_{kj}$ and four dissipative functions $r^S_{kj}$ of $\sigma$; $\mathcal S_{R^2}$ contributes only conservative $c^S_{10},c^S_{11}$.}
{We extract the angle from the recoil-corrected relative impulse $\Delta\vec p=\Delta\vec p_1+(E_1/E)\vec P_{\rm rad}$, with body-1 impulse $\Delta\vec p_1$, incoming center-of-mass energy $E_1$, and radiated linear momentum $\vec P_{\rm rad}$}~\cite{Dlapa:2022lmu,Jakobsen:2023hig}.
{Body exchange ($\delta\to-\delta$) relates the body-2 self-spin, quadrupole, and curvature-squared coefficients to those of body 1; odd-$\delta$ coefficients vanish for $\ah_1\ah_2$.}

{The spinning-test-body limits, in which one body is much lighter than the other, fix the $\nu$-independent combinations $\Theta^\pm_S=c^S_{00}\pm c^S_{01}$, where $\delta\to\pm1$ makes body~1 or~2 light. The probe may carry a spin-induced quadrupole.}
{The explicit probe-limit coefficients are given in~Eq.~\eqref{eq:probe}.} The genuinely two-body information resides in $c^S_{1j}$ and $r^S_{kj}$.

All functions for $S\in\mathcal S_0$ in Eq.~\eqref{eq:mass} are obtained in closed form, in the same function space as the spinless and spin-orbit
angles~\cite{Bern:2021yeh,Dlapa:2022lmu,Jakobsen:2023ndj,Jakobsen:2023hig}. With $w=(1-x)/(1+x)$, $z=w^2$,
$h=\operatorname{arccosh}\sigma$ and $L_\pm=\log(1\pm z)$,
each function takes the form
\begin{equation}
\pi\sum_{p,t=0}^{1}\sum_{n=1}^{14} w^p R_{pnt}(z)F_n(w)(\log w)^t ,
\label{eq:closed}
\end{equation}
with rational functions $R_{pnt}$ specific to the function and the fourteen basis functions
\begin{gather*}
\begin{alignedat}{3}
F_1&=1,     &\quad F_2&=h/w,      &\quad F_3&=L_-,\\
F_4&=L_+,   & F_5&=h^2,           & F_6&=hL_-/w,\\
F_7&=L_-^2, & F_8&=\operatorname{Li}_2(z),
                              & F_9&=\operatorname{Li}_2(-z),
\end{alignedat}\\
F_{10}=\frac{\operatorname{Li}_2(w)-\operatorname{Li}_2(-w)}{w},\\
\begin{alignedat}{2}
F_{11}&=\pi^2,  &\qquad F_{12}&=K(z)^2,\\
F_{13}&=E(z)^2, & F_{14}&=K(z)E(z).
\end{alignedat}
\end{gather*}
Here $K$ and $E$ are complete elliptic integrals of the first and
second kinds, respectively, with parameter $z$.
{No new transcendental functions arise beyond the nonspinning fourth-PM problem. Complete expressions are in \texttt{anc/}, with conventions and benchmarks in the Supplemental Material (SupMat)}~\cite{SM}.
{The curvature-squared coefficients, rational in $\sigma$ times $\pi$, are given explicitly in~Eq.~\eqref{eq:r2s2}.}

\begin{figure}[t]
\centering
\resizebox{\columnwidth}{!}{\begin{tikzpicture}[
  font=\fontsize{8}{9}\selectfont,
  bg/.style={densely dashed,line width=.5pt,black!55},
  zl/.style={line width=1.1pt},
  psl/.style={line width=1.1pt,postaction={decorate},decoration={markings,
    mark=at position .6 with {\arrow{Stealth[length=4pt,width=3.6pt]}}}},
  gr/.style={line width=.45pt,decorate,decoration={snake,amplitude=.75pt,segment length=3.4pt}},
  rad/.style={line width=.8pt,red!75!black,decorate,decoration={snake,amplitude=.9pt,segment length=3.4pt}},
  v/.style={circle,fill=black,inner sep=1.0pt},
  bulk/.style={circle,fill=black,inner sep=1.0pt},
  sv/.style={diamond,fill=black,inner sep=1.5pt},
  cut/.style={line width=1.1pt,-{Stealth[length=4.5pt,width=4pt]}}]
\begin{scope}
  \node[anchor=west] at (-.45,1.25) {(a) Both bodies respond};
  \node[anchor=north east,font=\fontsize{7.5}{8}\selectfont] at (3.2,-.68) {$m_1^2m_2^3$};
  \draw[bg] (-.3,.6)--(3.2,.6); \draw[bg] (-.3,-.6)--(3.2,-.6);
  \node[left] at (-.3,.6) {$1$}; \node[left] at (-.3,-.6) {$2$};
  \foreach \x in {0,.8,1.6,2.4} {\draw[gr] (\x,.6)--(\x,-.6); \node[v] at (\x,.6){}; \node[v] at (\x,-.6){};}
  \draw[zl] (0,-.6)--(.8,-.6);
  \draw[psl] (.8,.6)--(1.6,.6);
  \draw[zl] (1.6,.6)--(2.4,.6);
  \draw[cut] (2.4,.6)--(3.15,1.0);
\end{scope}
\begin{scope}[xshift=4.25cm]
  \node[anchor=west] at (-.45,1.25) {(b) Spin vertex};
  \node[anchor=north east,font=\fontsize{7.5}{8}\selectfont] at (3.2,-.68) {$m_1m_2^4$};
  \draw[bg] (-.3,.6)--(3.2,.6); \draw[bg] (-.3,-.6)--(3.2,-.6);
  \node[left] at (-.3,.6) {$1$}; \node[left] at (-.3,-.6) {$2$};
  \foreach \x in {0,.8,1.6,2.4} {\draw[gr] (2.1,.6)--(\x,-.6); \node[v] at (\x,-.6){};}
  \node[sv] at (2.1,.6){};
  \draw[cut] (2.1,.6)--(2.85,1.0);
\end{scope}
\begin{scope}[yshift=-2.5cm]
  \node[anchor=west] at (-.45,1.25) {(c) Self-interaction};
  \node[anchor=north east,font=\fontsize{7.5}{8}\selectfont] at (3.2,-.68) {$m_1^2m_2^3$};
  \draw[bg] (-.3,.6)--(3.2,.6); \draw[bg] (-.3,-.6)--(3.2,-.6);
  \node[left] at (-.3,.6) {$1$}; \node[left] at (-.3,-.6) {$2$};
  \coordinate (w) at (1.3,0);
  \draw[gr] (.9,.6)--(w); \draw[gr] (1.7,.6)--(w);
  \draw[gr] (w)--(.2,-.6); \draw[gr] (w)--(1.3,-.6); \draw[gr] (w)--(2.4,-.6);
  \foreach \p in {(.9,.6),(1.7,.6),(.2,-.6),(1.3,-.6),(2.4,-.6)} \node[v] at \p{};
  \node[bulk] at (w){};
  \draw[cut] (1.7,.6)--(2.45,1.0);
\end{scope}
\begin{scope}[xshift=4.25cm,yshift=-2.5cm]
  \node[anchor=west] at (-.45,1.25) {(d) Radiative graviton};
  \node[anchor=north east,font=\fontsize{7.5}{8}\selectfont] at (3.2,-.68) {$m_1^2m_2^3$};
  \draw[bg] (-.3,.6)--(3.2,.6); \draw[bg] (-.3,-.6)--(3.2,-.6);
  \node[left] at (-.3,.6) {$1$}; \node[left] at (-.3,-.6) {$2$};
  \coordinate (A) at (.45,0); \coordinate (B) at (1.55,0);
  \draw[gr] (0,.6)--(A); \draw[gr] (A)--(.45,-.6);
  \draw[rad] (A)--(B);
  \draw[gr] (B)--(1.55,-.6); \draw[gr] (B)--(1.85,.6);
  \draw[zl] (1.85,.6)--(2.5,.6);
  \draw[gr] (2.5,.6)--(2.5,-.6);
  \foreach \p in {(0,.6),(.45,-.6),(1.55,-.6),(1.85,.6),(2.5,.6),(2.5,-.6)} \node[v] at \p{};
  \node[bulk] at (A){}; \node[bulk] at (B){};
  \draw[cut] (2.5,.6)--(3.2,1.0);
\end{scope}
\end{tikzpicture}}
\caption{Representative $G^4$ worldline diagrams contributing to the impulse on body~1; mass factors label diagrams. Dashed lines are unperturbed trajectories, wavy lines are gravitons, and solid segments carry position (plain) or spin (arrow) responses; the outgoing arrow denotes the impulse. (a) Repeated exchanges involving the position response of body 2 and the spin and position responses of body 1. (b) Four gravitons attached to a spin-dependent worldline vertex (diamond). (c) Bulk graviton self-interaction. (d) A bulk--bulk graviton line (red) admits a radiative region, producing a conservative tail and radiation reaction. The $m_1m_2^4$ sector gives the probe limit.}
\label{fig:diagrams}
\end{figure}
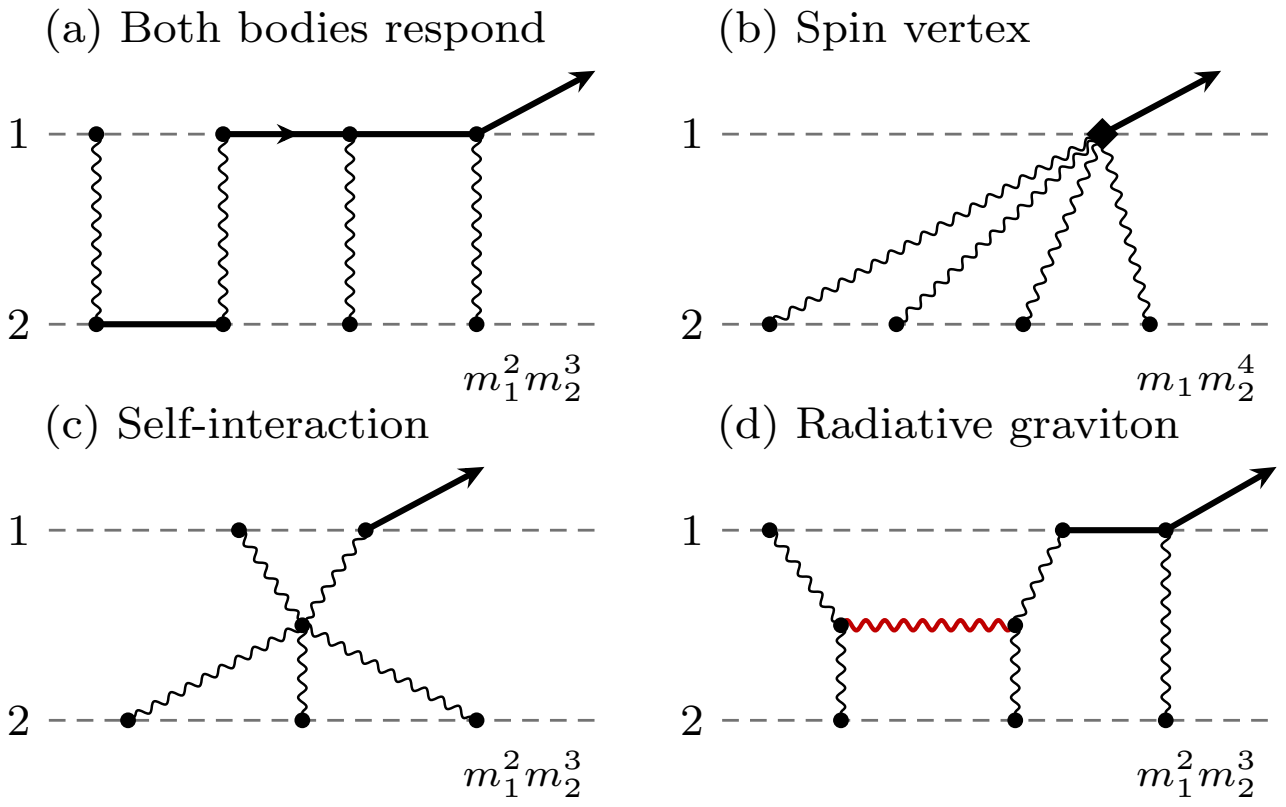

\sechead{Method}---We describe each body by the supersymmetric spinning worldline of
Ref.~\cite{Jakobsen:2021zvh}, which is exact through quadratic order in spin and contains the spin-induced
quadrupole, and we compute the impulse with worldline quantum field
theory~\cite{Mogull:2020sak,Jakobsen:2021smu}. In this formalism the
classical solution is a sum of tree graphs: gravitons are
exchanged between the two paths and interact among
themselves, while each body responds through changes
of its position and spin (Fig.~\ref{fig:diagrams}). Retarded propagators
make the response causal, so that radiation reaction is
included from the outset~\cite{Jakobsen:2022fcj}. {At fourth order there are 2620 graphs; the measured impulse
is obtained after including the recoil of the radiating
two-body system.}

{Worldline time integration yields three-loop integrals with delta constraints, linear worldline denominators, and quadratic graviton propagators. Their causal prescriptions are
resolved into principal-value and cut contributions,
retaining the energy-sign weights of radiative cuts (SupMat~\cite{SM}). Spin raises numerator ranks without introducing new integral sectors}~\cite{Dlapa:2021npj,Jakobsen:2023ndj}.
We build the numerators with FORM~\cite{Vermaseren:2000nd,Ruijl:2017dtg} and reduce about $2\times10^5$
integrals to master integrals by integration by parts~\cite{Chetyrkin:1981qh,Laporta:2000dsw}, using
Kira~\cite{Klappert:2020nbg,Lange:2025fba} and FireFly~\cite{Klappert:2019emp,Klappert:2020aqs}. {After factoring out overall momentum transfer $q^2$, the master integrals depend only on $x$ (or $\sigma$); their differential equations}~\cite{Henn:2013pwa} {are solved to about 40 digits using static boundary data from expansion by regions}~\cite{Beneke:1997zp}. {One sector is elliptic}~\cite{Bern:2021dqo,Dlapa:2021npj}; see also Ref.~\cite{Ahmed:2024tsg}
for related differential-equation techniques. {The potential and time-symmetric part of the two-radiative-graviton region give the conservative angle; the other radiative terms give the dissipative one, retaining the causal and regulator dependence needed for cancellations.}

To obtain Eq.~\eqref{eq:closed}, we compute exact
series in $w$ through $w^{400}$ for the conservative
angle and the part quadratic in radiation losses,
and through $w^{112}$ for the part linear in the
losses. These series determine the
rational coefficients of the closed-form ansatz; the final
expressions are not truncated in $w$. The basis and
the allowed denominators are those of the differential
equations and of the lower-spin angles. The linear
systems are overdetermined: 3499 coefficients are fixed
uniquely within this function space by 8467 exact
equations, with all 4968 surplus conditions satisfied
identically. Direct
numerical evaluations agree to at least 35 digits at eight
velocities and five mass ratios~\cite{SM}.

{For $\mathcal S_{R^2}$, inserting two linearized companion curvatures on straight worldlines yields one-loop, two-graviton exchange at $G^4S^2$. The impulse gives purely conservative coefficients $c^S_{10},c^S_{11}$, with no radiation reaction or tail logarithms. See Eq.~\eqref{eq:r2s2op} and~\EM{E} for the operators and SupMat~\cite{SM} for independent checks.}

\sechead{Infrared structure and tail}---{The gravitational-wave tail, radiation scattered by the binary's long-range field, first affects conservative scattering at $G^4$}~\cite{Blanchet:1987wq,Bern:2021yeh}. {In the three-loop calculation, infrared poles from the potential and two-radiative-graviton regions cancel in $d=4-2\epsilon$, leaving a velocity-dependent logarithm.}

We separate the three-loop conservative
angle into a potential-region contribution $\theta_{\rm pot}$ and a
time-symmetric tail $\theta_{\rm tail}$, and the dissipative angle into
contributions linear and quadratic in radiation reaction,
$\theta_{\rm lin},\theta_{\rm quad}$ (not spin degree). 
If $\mathcal P$ denotes the coefficient of $\log[(\sigma-1)/2]$ in the fourth-order angle,
including its PM prefactor, the tail is governed by the energy radiated at the preceding
order~\cite{Bern:2021dqo,Bern:2021yeh,Jakobsen:2023ndj},
\begin{equation}
\mathcal P=\frac{GE}{\pinf}\,\frac{\partial E^{(3)}_{\rm rad}}{\partial b}\bigg|_{E,a_i} .
\label{eq:tail}
\end{equation}
A term of spin degree $k$ in $E^{(3)}_{\rm rad}$ scales as $b^{-3-k}$ and contributes
$-(3+k)GE\,E^{(3)}_{{\rm rad},k}/(\pinf b)$. Using the radiated energy through quadratic order in spin of
Ref.~\cite{Riva:2022fru}, which our own $\mathcal{O}(G^3)$ calculation reproduces, we find the exact identities
\begin{equation}
\begin{aligned}
\theta_{\rm pot}&=\frac{\mathcal P}{2\epsilon}+\mathcal O(\epsilon^0),\qquad
\theta_{\rm tail}=-\frac{\mathcal P}{2\epsilon}+\mathcal O(\epsilon^0),\\
\theta_{\rm lin}&=\mathcal O(\epsilon^0),\qquad \theta_{\rm quad}=\mathcal O(\epsilon^0){\color{newtextcolor}.}
\end{aligned}
\label{eq:poles}
\end{equation}
{The identities hold for all velocities and separately for each mass-ratio coefficient of all eight $S\in\mathcal S_0$; all higher poles cancel. The curvature-squared sector $\mathcal S_{R^2}$ is infrared finite, with no tail logarithms,} leaving
these identities unchanged.
{The tail relation fixes the logarithm PN coefficient by low-velocity expansion. Elliptic integrals cancel algebraically from the poles~\EM{B}.}

\begin{figure}[t]
\includegraphics[width=\columnwidth]{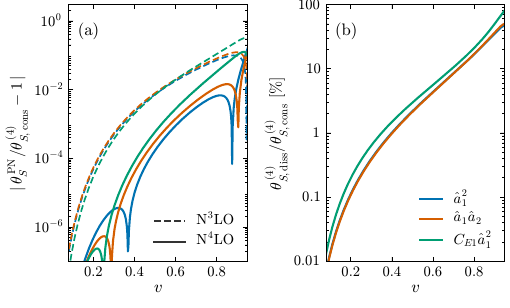}
\caption{Quadratic-spin coefficients of $\theta^{(4)}$ for equal masses versus relative speed $v$,
for $\ah_1^2$, $\ah_1\ah_2$, and $\CE{1}\ah_1^2$ (colors identified in panel~(b)).
(a) Absolute relative PN deviation $|\theta^{(4)}_{S,\mathrm{PN}}/\theta^{(4)}_{S,\mathrm{cons}}-1|$,
with the PN series of $\theta^{(4)}_{S,\mathrm{cons}}/\Gamma$ truncated at the previously known N$^3$LO terms (dashed) and including the new N$^4$LO term (solid).
(b) Dissipative-to-conservative ratio $\theta^{(4)}_{S,\mathrm{diss}}/\theta^{(4)}_{S,\mathrm{cons}}$ in percent.
Both panels show individual spin coefficients, not the total angle.}
\label{fig:result}
\end{figure}

\sechead{Post-Newtonian and velocity dependence}---
{At low velocity, the five quadratic-spin conservative structures in $\mathcal S_0$ begin at $v^{-6}$; their $v^{-6},v^{-4},v^{-2}$ terms agree with N$^3$LO PN results}~\cite{Kim:2022bwv,Mandal:2022ufb}. {The $v^0$ term is new.}
{It is $f_{0,S}+t_{0,S}\log[(\sigma-1)/2]$ in $\theta^{(4)}_{S,\mathrm{cons}}/\Gamma$, the first logarithmic order. At $v=3/5$ it reduces the displayed PN deviation from about $2\%$ to $0.1$--$0.3\%$ [Fig.~\ref{fig:result}(a)].} For the mixed-spin structure,
\begin{equation}
\begin{aligned}
f_{0,\ah_1\ah_2}&=\frac{525\pi}{16}+\pi\nu\left(-\frac{17455}{12}+\frac{641305\pi^2}{16384}\right),\\
t_{0,\ah_1\ah_2}&=-\frac{227\pi\nu}{4} .
\end{aligned}
\label{eq:pn}
\end{equation}
{Other PN coefficients follow to arbitrary order from \texttt{anc/}; selected terms are given in SupMat}~\cite{SM}.

{In $\mathcal S_{R^2}$ the leading electric term, at $v^{-2}$, involves $6C^{(2)}_{E^2,i}+C^{(0)}_{E^2S^2,i}$, agreeing with} Refs.~\cite{Kim:2022bwv,Mandal:2022ufb}.
At $\mathcal O(v^0)$, both electric and magnetic
couplings contribute through
\begin{align}
\mathcal Q_i={}&
9C^{(2)}_{E^2,i}
-C^{(0)}_{E^2S^2,i}
+\frac52\big(
6C^{(2)}_{B^2,i}
-C^{(0)}_{B^2S^2,i}\big),
\end{align}
giving
\begin{align}
\left.\frac{\theta^{(4)}}{\Gamma}\right|_{v^0}
\supset \frac{15\pi\nu}{32}\Big[
&(1-\delta)\ah_1^2\mathcal Q_1
+(1+\delta)\ah_2^2\mathcal Q_2\Big].
\label{eq:r2pn}
\end{align}
These contributions contain no logarithms.

\begin{figure*}[t]
\includegraphics[width=\textwidth]{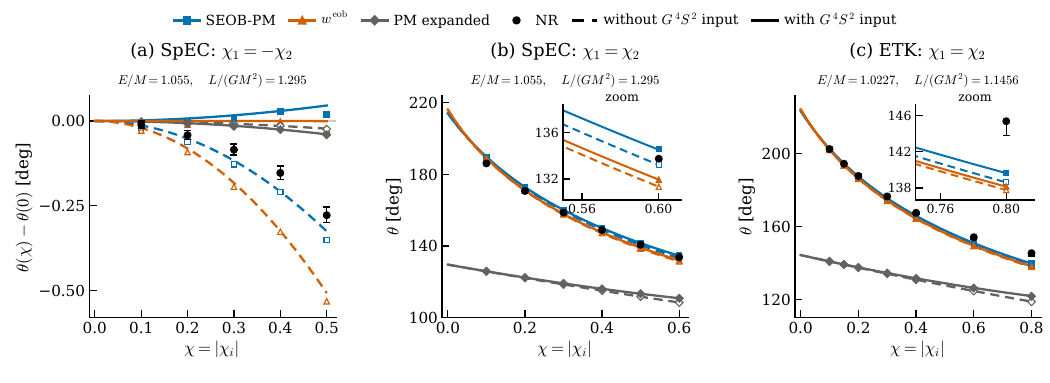}
\caption{Effect of the computed $G^4S^2$ input on equal-mass black-hole scattering:
(a) opposite-spin difference $\theta(\chi,-\chi)-\theta(0,0)$ in SpEC~\cite{Long:2025nmj},
(b) equal-spin angle in SpEC~\cite{Long:2025nmj}, and (c) equal-spin angle in ETK~\cite{Rettegno:2023ghr}.
Blue, orange, and gray curves denote SEOB-PM, $w^{\rm eob}$, and the PM-expanded prediction.
Dashed (solid) curves exclude (include) the new $G^4S^2$ input, with all other inputs fixed.
Smooth curves use the displayed nominal $E/M$ and canonical $L/(GM^2)$; open (filled) colored markers use
individual runs' exact initial conditions without (with) the new input.
Black points are NR results with asymmetric uncertainties, propagated through the subtraction in~(a).
Insets enlarge the high-spin region.}
\label{fig:nr}
\end{figure*}
The size of dissipative effects is not negligible at relativistic velocities: for equal masses, radiation reaction contributes $4$--$6\%$ of the conservative quadratic-spin coefficients {in $\mathcal S_0$} at $v=3/5$, rising to $35$--$50\%$ at $v=0.9$; the contribution quadratic in losses stays below $0.03\%$ at $v=3/5$ [Fig.~\ref{fig:result}(b)].

\sechead{Radiated angular momentum}---To first order in radiated losses, the dissipative scattering angle is fixed by the response of the conservative angle to the radiated energy and angular
momentum~\cite{Bini:2012ji,Damour:2020tta,Manohar:2022dea,Jakobsen:2022zsx},
\begin{equation}
\theta_{\rm rad,1}=-\frac12\Big(\frac{\partial\theta}{\partial J}J_{\rm rad}+\frac{\partial\theta}{\partial E}E_{\rm rad}\Big) {\color{newtextcolor}.}
\label{eq:linresp}
\end{equation}
{Derivatives hold physical spins fixed; $\partial/\partial E$ also holds total, not orbital, $J$ fixed.}
{At $G^4$ this involves $J_{\rm rad}^{(3)}$, known without spin}~\cite{Manohar:2022dea}, {at linear spin}~\cite{Jakobsen:2023hig,Heissenberg:2023uvo}, {and at quadratic spin for black holes}~\cite{Zhu:ZenodoKerr}.
For aligned spins we write~\cite{Jakobsen:2023hig}
\begin{equation}
J^{(3)}_{\rm rad}=-\big[E^{(3)}_{\rm rad}J+\pi G^3m_1^2m_2^2X/b^2\big]/E ,
\label{eq:jrad}
\end{equation}
{where the second term gives the change in the product $EJ$} of the two bodies~\cite{Jakobsen:2026ayw}. {For the five quadratic-spin structures $S\in\mathcal S_0$, we write $X\supset M\sum_S X_S S$ with $X_S=X_S^{(0)}+\delta X_S^{(1)}$. The known $E_{\rm rad}^{(3)}$ through quadratic spin}~\cite{Riva:2022fru} {is reproduced by our $G^3$ calculation. Curvature-squared operators first enter scattering at $G^4$ and do not contribute here.}

We determine $X_S$ independently from the fourth-PM
linear-in-loss angle using Eqs.~\eqref{eq:linresp}
and~\eqref{eq:jrad}, and from a direct $G^3$
worldline calculation of the impact-parameter shift.
The latter involves differentiating the outgoing-line
current with respect to its frequency, together with
a zero-frequency radiation contribution~\cite{Jakobsen:2026ayw}.
The two determinations agree to 39 digits at four
velocities and four mass ratios for all five spin
structures. The resulting closed forms~\EM{C}
have static limits independently reproduced from
mass- and current-quadrupole radiation~\cite{Kidder:1995zr,Gergely:2002fd}
along straight-line trajectories perturbed by
leading spin forces.
These checks validate the linear-in-loss angle,
but not its finite quadratic-in-loss contribution.
For Kerr black holes, our third-PM angular-momentum
loss and fourth-PM linear-in-loss angle agree,
after restricting to aligned spins and matching
conventions, with the generic-spin results of
Ref.~\cite{Zhu:ZenodoKerr}. 
{Thus we obtain closed-form third-PM angular-momentum loss through quadratic spin for general spin-induced quadrupoles, complementing recent nonspinning fourth-PM results}~\cite{Heissenberg:2025ocy,Georgoudis:2026jlo,Bohnenblust:2026mal}.

\sechead{Checks}---{The spinless and linear-spin limits reproduce published conservative and dissipative angles as exact identities, testing the integrands, reductions, and master integrals used for quadratic spin}~\cite{Bern:2021yeh,Dlapa:2022lmu,Jakobsen:2023ndj,Jakobsen:2023hig}. {The three-loop quadratic-spin terms reproduce probe limits, satisfy infrared-pole and linear-response identities using independently computed $J_{\rm rad}^{(3)}$, agree with known PN orders, and pass numerator, reduction, and on-shell checks (\EM{A}; SupMat~\cite{SM}). Curvature-squared terms also agree with PN results.}

\sechead{Comparison with numerical relativity}---EOB resummation reorganizes
perturbative gravitational dynamics into an effective
one-body description, extending its predictive reach
to strong-field black-hole scattering. {We insert the computed $G^4S^2$ Kerr coefficient into two EOB resummations, SEOB-PM}~\cite{Buonanno:2024vkx} {and $w^{\rm eob}$}~\cite{Damour:2022ybd,Rettegno:2023ghr}, {and compare with equal-mass black-hole scattering simulations from the Spectral Einstein Code (SpEC)}~\cite{Long:2025nmj} {and the Einstein Toolkit (ETK)}~\cite{Rettegno:2023ghr} {[Fig.~\ref{fig:nr}], where curvature-squared couplings vanish. For SpEC we use the mean of the two black-hole angles}~\cite{Long:2025nmj}. {These are strongly deflected trajectories: the nonspinning SpEC angle is $207.0^\circ$, versus $129.5^\circ$ from the PM expansion, requiring resummation. At each run's energy we replace the model-implied $G^4S^2$ coefficient by ours, fixing all known $G^4S^1$, $G^3S^2$, and higher-spin inputs~\EM{D}; only this replacement distinguishes dashed and solid curves.}

{At equal SpEC spins $\chi\simeq0.6$, SEOB-PM shifts from $133.24^\circ$ to $134.56^\circ$ and $w^{\rm eob}$ from $131.33^\circ$ to $131.95^\circ$, versus NR's $133.74^\circ\pm0.02^\circ$. The SEOB-PM shift of $1.31^\circ$ is about 70 times the NR uncertainty but overshoots it. At ETK $\chi=0.8$, $w^{\rm eob}$ shifts from $137.79^\circ$ to $138.17^\circ$, toward NR's $145.36^\circ$ (asymmetric uncertainty up to $1.5^\circ$). Opposite equal-magnitude spins cancel odd-spin terms at equal masses; subtracting the nonspinning angle isolates even-spin effects. At $\chi=0.5$ NR gives $-0.28^\circ$, while SEOB-PM changes from $-0.35^\circ$ to $+0.02^\circ$ and $w^{\rm eob}$ from $-0.53^\circ$ to $-0.03^\circ$.}

The input-induced shift exceeds SpEC's quoted uncertainty
by over an order of magnitude for equal spins and, for
opposite spins, from $\chi=0.4$.
Yet agreement with NR remains model dependent:
at equal spins $w^{\rm eob}$ moves toward NR except
for the two smallest SpEC spins, while SEOB-PM moves
away for $\chi\leq0.4$ and crosses NR for $\chi\geq0.5$.
In ETK, SEOB-PM moves toward NR, crossing it near
$\chi=0.1$; for opposite spins, NR lies between
predictions with and without our input.
For equal spins, the residual discrepancies are
comparable to or larger than the new contribution.
In particular, for opposite spins with $\chi\ge0.2$ the updated
SEOB-PM prediction moves farther from NR,
illustrating that incorporating an exact weak-field
coefficient need not improve a strong-field resummation.
Thus these resolvable, model-dependent shifts do not
independently determine the weak-field $G^4S^2$
coefficient. Our result provides an exact perturbative
constraint for testing strong-field resummations.

\sechead{Conclusions}---{We have determined the fourth-PM aligned-spin scattering angle through quadratic spin, including spin-induced quadrupoles, conservative curvature-squared spin couplings, and radiation reaction. Closed forms, exact cancellation of potential--tail infrared poles, new PN information, and independently determined third-PM angular-momentum loss provide complementary predictions and checks. The new Kerr-sector coefficient changes EOB-resummed predictions appreciably, while agreement with NR remains model dependent. Applications to bound-orbit dynamics and waveforms, generic spin orientations, and spin kicks remain open.}

\sechead{Acknowledgments}---
We thank Manoj K. Mandal for comments on the draft and for valuable suggestions.
The work of S.M.H. is supported by the Academic Research Sabbatical Program (ARSP) of the University of Regensburg.
The authors gratefully acknowledge the computational and data resources provided by the Leibniz Supercomputing Centre (www.lrz.de).

\sechead{Data availability}---The exact closed forms and evaluation files are provided in \texttt{anc/}, with conventions, analytic checks, and representative benchmarks in SupMat~\cite{SM}.

\bibliography{refs}

\raggedbottom
\onecolumngrid
\section*{End Matter}
\twocolumngrid

\emhead{A}{Checks of the quadratic-spin terms}
{The $\nu$-independent conservative coefficients satisfy $\Theta^\pm_S=c^S_{00}\pm c^S_{01}$. With $s=\sigma$, $R=(s^2-1)^3$, their exact probe-limit functions are}
\begin{flalign}
&\Theta^+_{\ah_1^2}=\tfrac{75\pi}{32R}(231s^6-294s^4+91s^2-4), &&\nonumber\\
&\Theta^-_{\ah_1^2}=\tfrac{105\pi}{64R}(759s^6-1005s^4+325s^2-15), &&\nonumber\\
&\Theta^+_{\CE{1}\ah_1^2}=-\tfrac{75\pi}{64R}(231s^6-273s^4+77s^2-3), &&\nonumber\\
&\Theta^-_{\CE{1}\ah_1^2}=-\tfrac{15\pi}{896R}\!(29225s^6\!-\!40375s^4\!+\!15115s^2\!-\!1277), &&\nonumber\\
&\Theta^\pm_{\ah_1\ah_2}=\tfrac{105\pi}{16R}(264s^6-345s^4+110s^2-5). &&\label{eq:probe}
\end{flalign}

The lower-spin reproduction described in the main text is exact: the closed forms agree with the published
angles as identities between rational coefficients, and every master-integral value, region and sign entering
the quadratic-spin angle enters these lower orders with nonzero weight. The quadratic-spin terms are tested
further. (i) The probe limits in Eq.~\eqref{eq:probe} agree with the closed expressions, four of which also follow from the test-body
results of Ref.~\cite{Hoogeveen:2025tew}. (ii) The three known PN orders {were checked against
Refs.~\cite{Kim:2022bwv,Mandal:2022ufb}}. (iii) The pole identities~\eqref{eq:poles} hold with the independently known energy
loss. (iv) The radiated angular momentum extracted from the angle agrees with its direct calculation to {39} digits.
(v) The conservative impulse is symmetric under exchange of the bodies, mass sector by mass sector, although
different graphs contribute. (vi) The component of the impulse along the velocity agrees with its on-shell
prediction from lower orders~\cite{Jakobsen:2021zvh,Jakobsen:2022fcj}; this tests 888 of the 898
potential-region graphs, and the other ten are probe graphs covered by (i). (vii) A sample of 429 reduction
relations of high numerator rank is reproduced with a different integral ordering. (viii) A separately written
assembly code, using the same integral coefficients and master integrals, agrees to $10^{-38}$. The finite tail and the part quadratic in losses are not separately fixed by an independent observable beyond these tests; they rely on the shared integrands, reductions, and master integrals validated above.

\emhead{B}{Infrared poles}
The residues of Eq.~\eqref{eq:poles} are $\pi$ times rational functions of $x$ multiplying $1$, $\log x$ and
$\log[(1+x)/2]$. In the potential region the elliptic sector drops out algebraically: wherever it
could reach the pole, the integrand is a total derivative of a rational combination of the sector's own
functions, and the coefficients sum to zero. In linear radiation reaction the pole cancels between lower and
upper one-graviton master integrals and between the single-graviton cuts of the two integral families, with the
boundary constant involved kept symbolic; in quadratic radiation reaction it cancels between the two weightings
of the double cut. None of these cancellations uses numerical input.

\emhead{C}{Radiated angular momentum}
With $s\equiv\sigma$, the direct calculation gives for the mixed-spin structure
\begin{flalign}
&X_{\ah_1\ah_2}=\frac{3s(2s^2-3)p_5}{(s^2-1)^{5/2}}\,{\rm arccosh}\,s&&\nonumber\\
&\qquad-\frac{3p_6}{2(s-1)(s+1)^2}\log\frac{s+1}{2}-\frac{p_7}{8(s^2-1)^2} ,&&
\label{eq:x2}
\end{flalign}
with
\begin{flalign*}
&p_5=7s^5-40s^4-14s^3+42s^2+5s-6 ,&&\\
&p_6=28s^6-97s^5-181s^4-206s^3+610s^2&&\\
&\qquad-225s+23 ,&&\\
&p_7=168s^7-1619s^6+319s^5+5050s^4-7794s^3&&\\
&\qquad+8873s^2-6229s+1376 ,&&
\end{flalign*}
and $X^{(1)}_{\ah_1\ah_2}=0$. The self-spin and quadrupole structures have the same form and are given in
SupMat~\cite{SM}. In the static limit $s\to1$ the functions odd in $\delta$ vanish and
$X_{\ah_1^2}\to-51/4$, $X_{\ah_1\ah_2}\to-249/10$, $X_{\CE{1}\ah_1^2}\to63/5$; the elementary multipole
calculation reproduces all three values exactly.

\emhead{D}{Numerical-relativity inputs}
The fourth-order input is evaluated directly at the reported energy of each of the 60 SpEC and 35 ETK runs,
with Kerr values $C_{E,i}=0$ and the total angle including radiation reaction. Reported decimal energies and mass
ratios are used as exact rationals, with no interpolation in energy. Initial data specify the canonical orbital
angular momentum $L_{\rm can}=J-m_1a_1-m_2a_2$, which is related to the covariant impact parameter by
\begin{equation}
b=\frac{L_{\rm can}-(E_1-m_1)a_1-(E_2-m_2)a_2}{\pinf} ,
\end{equation}
with $E_i$ the energies of the bodies in the center-of-mass frame. In each resummation the $G^4S^2$ coefficient
implied by the model is replaced by the computed one, with all other spin inputs kept fixed. For SpEC the
error intervals of the mean angle are propagated without assuming independence of the two black-hole angles.
Model references are given in the Letter; values for the plotted comparisons are included in \texttt{anc/nr\_comparison.txt}.

\emhead{E}{Curvature-squared finite-size terms}
Generic compact bodies admit additional worldline
couplings quadratic in the
curvature~\cite{Mandal:2023lgy,Mandal:2023hqa}.
Through quadratic order in spin, four parity-even
operators contribute at fourth PM
order~\cite{Kim:2021rfj,Kim:2022bwv,Mandal:2022ufb},
\begin{align}
\mathcal L_i\supset\frac{G^2m_i}{2}\Big[
&C^{(2)}_{E^2,i}S_i^2E_{\mu\nu}E^{\mu\nu}
\nonumber\\
&+C^{(0)}_{E^2S^2,i}
E_{\mu\alpha}E_\nu{}^{\alpha}
(S_i^\mu S_i^\nu)_{\rm STF}
+(E\to B)\Big],
\label{eq:r2s2op}
\end{align}
where
$E_{\mu\nu}=R_{\mu\alpha\nu\beta}u_i^\alpha u_i^\beta$,
$B_{\mu\nu}$ is its magnetic counterpart,
$S_i^2=-S_{i\mu}S_i^\mu$ (mostly-minus signature),
and STF denotes the symmetric trace-free projection
orthogonal to $u_i^\mu$.

{For $S=C_{X,i}\ah_i^2$, the only nonzero fourth-PM coefficients of Eq.~\eqref{eq:mass} are}
\begin{align}
(c_{10}^{E2,i},c_{10}^{ES2,i})
&=\frac{15\pi}{512(\sigma^2-1)}
(6p_E,-q_E),
\nonumber\\
(c_{10}^{B2,i},c_{10}^{BS2,i})
&=\frac{75\pi(7\sigma^2+1)}{512}
(6,-1)
\label{eq:r2s2}
\end{align}
with
$p_E=35\sigma^4-30\sigma^2+11$,
$q_E=35\sigma^4-70\sigma^2+19$, and
$c_{11}^S=(-1)^i c_{10}^S$ for $i=1,2$. 
For brevity, the superscripts $E2$, $ES2$, $B2$, and $BS2$
on $c_{10}$ label the structures
$C^{(2)}_{E^2,i}\ah_i^2$,
$C^{(0)}_{E^2S^2,i}\ah_i^2$,
$C^{(2)}_{B^2,i}\ah_i^2$, and
$C^{(0)}_{B^2S^2,i}\ah_i^2$, respectively.

The four spin-tidal coefficients characterize the
spin-dependent tidal response of each body.
They vanish for Kerr black holes in the static
conservative matching~\cite{Chia:2020yla,
Charalambous:2021mea,Saketh:2023bul,Haddad:2024ebn}
and remain independent finite-size parameters
for generic compact objects.

Each operator carries an explicit factor $G^2$.
At fourth PM order, its contribution is obtained
from the two linearized curvatures sourced by the
companion, evaluated on straight trajectories.
The resulting correction is conservative and
free of logarithms. Since the operators are
quadratic in the spin of a single body, no
mixed-spin structure $\ah_1\ah_2$ arises at
this order. The corresponding coefficients
are given in Eq.~\eqref{eq:r2s2}.

\clearpage
\onecolumngrid
\setcounter{section}{0}
\setcounter{equation}{0}
\setcounter{table}{0}
\renewcommand{\thesection}{S\arabic{section}}
\renewcommand{\theequation}{S\arabic{equation}}
\renewcommand{\thetable}{S\arabic{table}}
\renewcommand{\theHsection}{SupMat.\arabic{section}}
\renewcommand{\theHequation}{SupMat.\arabic{equation}}
\renewcommand{\theHtable}{SupMat.\arabic{table}}
\bigskip

\begin{center}
{\large\bfseries Supplemental Material}\\[0.6ex]
{\normalsize\bfseries Quadratic-in-spin scattering of compact bodies at fourth post-Minkowskian order}\\[1ex]
Taushif Ahmed and Syed Mehedi Hasan\\
Institute for Theoretical Physics, University of Regensburg, Germany
\end{center}
\bigskip
\section{Conventions and ancillary files}
We use the kinematic, spin and mass conventions of the
Letter, in particular
$\sigma=(1+x^2)/(2x)$,
$w=(1-x)/(1+x)$,
$\delta=(m_2-m_1)/M$, and $C_{E,i}=0$ for Kerr
black holes.
The eight functions of Eq.~\eqref{eq:mass} are labelled
\texttt{A, B, C, D, E, F, G, H} in the ancillary files,
corresponding to
$c_{00}$, $c_{01}$, $c_{10}$, $c_{11}$,
$r_{00}$, $r_{01}$, $r_{10}$, $r_{11}$.
The curvature-squared structures $S\in\mathcal S_{R^2}$
use the same labels; only \texttt{C}~$=c^S_{10}$ and
\texttt{D}~$=c^S_{11}=(-1)^ic^S_{10}$ are nonzero
[Eq.~\eqref{eq:r2s2}]. Their names
\texttt{CEE}$i$, \texttt{CEES}$i$, \texttt{CBB}$i$ and
\texttt{CBBS}$i$ stand for
$C^{(2)}_{E^2,i}\ah_i^2$, $C^{(0)}_{E^2S^2,i}\ah_i^2$,
$C^{(2)}_{B^2,i}\ah_i^2$ and $C^{(0)}_{B^2S^2,i}\ah_i^2$,
the superscripts $E2$, $ES2$, $B2$ and $BS2$ of
Eq.~\eqref{eq:r2s2}.
For the radiated angular momentum, \texttt{A} and
\texttt{B} are $X^{(0)}_S$ and $X^{(1)}_S$ of
$X_S=X^{(0)}_S+\delta X^{(1)}_S$ in Eq.~\eqref{eq:jrad}.
The total result is the sum of conservative and
linear- and quadratic-in-loss contributions;
the linear-in-loss part includes the longitudinal
response to the radiated energy.
The complete elliptic integrals $K(z)$ and $E(z)$ use the
parameter $z$ (not the modulus $\sqrt z$).

The accompanying ancillary files are distributed in the \texttt{anc/} directory. They provide the complete fourth post-Minkowskian scattering-angle coefficients through quadratic order in spin as closed analytic functions of the relative velocity and mass ratio, without truncation in the velocity expansion. They contain the final analytic expressions and numerical comparisons, rather than the original graph-generation, integral-reduction, or master-integral production codes:
\begin{itemize}
\item \texttt{anc/closed\_forms.py} and \texttt{anc/ClosedForms.m}: exact closed forms of the functions $c^S_{kj}$ and $r^S_{kj}$ of Eq.~\eqref{eq:mass} for the eight spin structures $S\in\mathcal S_0$, in the form of Eq.~\eqref{eq:closed}, resolved into conservative, linear and quadratic radiation-reaction, dissipative, and total parts; their expansions give the probe limits of Eq.~\eqref{eq:probe} and the post-Newtonian coefficients of Eq.~\eqref{eq:pn};
\item \texttt{anc/radiated\_angular\_momentum.py} and \texttt{anc/RadiatedAngularMomentum.m}: the closed quadratic-spin functions $X_S$ of the third-order radiated angular momentum, Eq.~\eqref{eq:jrad}, which enter the linear response of Eq.~\eqref{eq:linresp}; $X_{\ah_1\ah_2}$ is given in Eq.~\eqref{eq:x2}, the other four structures in the section on the radiated angular momentum below;
\item {\texttt{anc/curvature\_squared.py} and \texttt{anc/CurvatureSquared.m}: the coefficients $c^S_{10}$ and $c^S_{11}$ of Eq.~\eqref{eq:mass} for the eight curvature-squared structures $S\in\mathcal S_{R^2}$, given in Eq.~\eqref{eq:r2s2} for the couplings of Eq.~\eqref{eq:r2s2op}, which vanish for Kerr black holes; their expansion gives Eq.~\eqref{eq:r2pn};}
\item \texttt{anc/evaluate\_complete\_angle.py}: a worked Python example of Eqs.~\eqref{eq:theta} and~\eqref{eq:mass}, combining the eight spin structures of $\mathcal S_0$, restoring $\Gamma$, and evaluating the physical fourth-order contribution {with vanishing curvature-squared couplings; \texttt{anc/README.md} shows how to add them};
\item \texttt{anc/nr\_comparison.txt}: the numerical-relativity and model values corresponding to Fig.~\ref{fig:nr} of the Letter, with the inputs of End Matter~D;
\item \texttt{anc/README.md}: conventions, function interfaces, prerequisites, benchmark values, and usage instructions.
\end{itemize}
The Python evaluation files require \texttt{sympy} and \texttt{mpmath}; the Mathematica expressions require no separate coefficient tables. These files allow evaluation of the final results, but do not regenerate the complete three-loop calculation.

\section{Method and numerical precision}

The spinning-body description uses the $\mathcal N=2$
supersymmetric worldline and is exact through quadratic
order in spin, including the spin-induced quadrupole
coefficient $C_E$. The fourth-order graph set includes
up to three spin lines on one worldline and a vertex
coupling four gravitons; retarded propagators impose
the causal response.

A representative integral before reduction has the
schematic form
\begin{equation}
I_{\boldsymbol a}\sim
\int\prod_{r=1}^{3}\frac{d^d\ell_r}{(2\pi)^d}\,
\frac{
\mathcal N(\{\ell\},q)
\prod_{i=1}^{3}\widehat\delta(u_{\alpha_i}\cdot L_i)
}{
\prod_j(u_{\beta_j}\cdot Q_j\pm i0)^{a_j}
\prod_k(K_k^2+i0\,\eta_k)^{b_k}
},
\label{eq:worldline_integral}
\end{equation}
where $q$ is the momentum transfer, $L_i,Q_j,K_k$
are linear combinations of loop momenta and $q$,
and $\widehat\delta(A)=2\pi\delta(A)$. The three
base worldline constraints can be chosen as
$u_1\cdot\ell_1=u_1\cdot\ell_2=u_2\cdot\ell_3=0$;
additional linear denominators involve
$u_2\cdot\ell_{1,2}$ and $u_1\cdot\ell_3$.
The signs $\pm$ distinguish advanced and retarded
worldline prescriptions. For a graviton square,
$\eta_k=+1$ denotes a Feynman prescription, while
$\eta_k=\pm\operatorname{sgn}(K_k^0)$ specifies
retarded or advanced propagation. For simple
retarded or advanced poles, the relevant
distributional identities are
\begin{equation}
\begin{aligned}
&\frac{1}{A\pm i0}
=\operatorname{PV}\frac{1}{A}
\mp i\pi\,\delta(A),\\[1ex]
&\frac{1}{K^2\pm i0\,\operatorname{sgn}(K^0)}
=\operatorname{PV}\frac{1}{K^2}
\mp i\pi\,\operatorname{sgn}(K^0)\,\delta(K^2).
\end{aligned}
\label{eq:cut_identities}
\end{equation}

Higher powers generate derivatives of delta functions.
The sign weight $\operatorname{sgn}(K^0)$ of a radiative
graviton cut must be retained during integration-by-parts
reduction; it cannot be treated as an unweighted on-shell
delta function. Different potential and radiation
contributions require the corresponding prescribed or
cut version of the same quadratic denominator.

The integrals belong to the four three-loop families
of the spinless calculation. The reduction distinguishes
integral families, kinematic regions, and cut prescriptions.
Each worldline propagator is separated into its
principal-value and delta-function parts.
References for the tools and methods are given in the
Letter. The integration-by-parts systems contain about
$10^9$ equations.
After factoring out the overall momentum-transfer
scale $q^2$, the three-loop master integrals depend
on a single nontrivial dimensionless kinematic
variable, $\sigma=u_1\cdot u_2$. Their differential
equations are solved in the rationalized variable $x$,
defined by $\sigma=(1+x^2)/(2x)$, with boundary
conditions at the static point $x=1$.
Their numerical evaluation carries about 40
significant digits. The region factors are
$(1-x)^{-2\epsilon}$ and $(1-x)^{-4\epsilon}$
for one and two radiative gravitons in
$d=4-2\epsilon$. The potential contribution
and the time-symmetric two-graviton contribution
give the conservative angle.

The closed forms are reconstructed within the function
basis inherited from the spinless and spin-orbit
fourth-order problems, and its allowed rational
denominators are fixed by the master differential
equations. This reconstruction is not a separate
derivation of the quadratic-spin integrand: the exact
rank and surplus tests establish uniqueness within
the specified finite-dimensional ansatz, but a finite
series alone cannot exclude functions outside that
basis. Independent finite-velocity production
comparisons and lower-spin analytical overlaps
provide additional support.
The 8467 equations and 3499 unknowns quoted in the
Letter are distributed over 64 linear systems, one for
each of the eight functions of the eight structures in
$\mathcal S_0$; the 4968 surplus equations are satisfied
with exactly zero residual.
At the four production velocities, $x=1/3,1/2,3/5,9/10$,
the stored (rounded) conservative and total angles
carry at least 30 significant digits, and the smaller
term quadratic in the losses between 24 and 33;
the comparisons with the direct evaluation in the
checks below use the unrounded production values.

\section{Post-Newtonian coefficients}
Expand the conservative functions at low velocity as $c^S_{kj}=\sum_n v^n c^S_{kj,n}$ plus logarithmic terms, and write the logarithmic part of the conservative angle as $(T_{4,S}/\Gamma)\log[(\sigma-1)/2]$. The orders $n=-6,-4,-2$ {were checked against} the published N$^3$LO post-Newtonian {results} cited in the Letter. The next order, $n=0$, is new:

\begin{table}[h]
\caption{Coefficients at $\mathcal{O}(v^0)$ of the conservative quadratic-spin angle. The structures $\ah_2^2$ and $C_{E,2}\ah_2^2$ follow from $\ah_1^2$ and $C_{E,1}\ah_1^2$ by $\delta\to-\delta$.}
\renewcommand{\arraystretch}{1.5}
\begin{ruledtabular}
\begin{tabular}{lccc}
 & $\ah_1^2$ & $\ah_1\ah_2$ & $C_{E,1}\ah_1^2$\\
\hline
$c^S_{00,0}$ & $\frac{2175\pi}{128}$ & $\frac{525\pi}{16}$ & $-\frac{22305\pi}{1792}$\\
$c^S_{01,0}$ & $-\frac{975\pi}{128}$ & $0$ & $\frac{16005\pi}{1792}$\\
$c^S_{10,0}$ & $-\frac{10096339\pi}{13440}+\frac{807625\pi^3}{32768}$ & $-\frac{17455\pi}{12}+\frac{641305\pi^3}{16384}$ & $\frac{315017\pi}{640}-\frac{1444945\pi^3}{65536}$\\
$c^S_{11,0}$ & $\frac{372067\pi}{2688}-\frac{136515\pi^3}{16384}$ & $0$ & $-\frac{338819\pi}{2688}+\frac{353035\pi^3}{32768}$\\
$[T_4/\Gamma]_{v^0}$ & $-\frac{245\pi\nu}{8}$ & $-\frac{227\pi\nu}{4}$ & $\frac{59\pi\nu}{2}$\\
\end{tabular}
\end{ruledtabular}
\end{table}

\section{Curvature-squared couplings}

The couplings of Eq.~\eqref{eq:r2s2op} of the Letter
use the mostly-minus signature, $S^2=-S_\mu S^\mu$,
$B_{\mu\nu}=\tfrac12\epsilon_{\alpha\beta\gamma\mu}
R^{\alpha\beta}{}_{\delta\nu}u^\gamma u^\delta$,
and the symmetric trace-free projection orthogonal to $u$.
The electric pair is normalized as in Eqs.~(2.10)
and (2.11) of Ref.~\cite{Mandal:2022ufb}, where
$C^{(2)}_{E^2}$ is the spin-squared part of the
tidal coefficient,
$C_{E^2}=C^{(0)}_{E^2}
+C^{(2)}_{E^2}S^2/(G^2m^4)$;
$C^{(0)}_{E^2S^2}$ and $C^{(0)}_{B^2S^2}$
equal $C_{E^2S^2}$ and $C_{B^2S^2}$ of
Eq.~(2.10b) of Ref.~\cite{Kim:2022bwv}, and
$C^{(2)}_{B^2}$ is defined like $C^{(2)}_{E^2}$.

In the rest frame of the body the couplings read
$\tfrac12G^2m\,[C^{(2)}_{E^2}S^2E_{ab}E_{ab}
-C^{(0)}_{E^2S^2}E_{ac}E_{bc}
(S^aS^b)_{\rm STF}+(E\to B)]$.
In the notation of Ref.~\cite{Haddad:2024ebn},
read with the same signature,
$C^{(1)}_{(E/B)^2}=
\tfrac12C^{(2)}_{(E/B)^2}
+\tfrac16C^{(0)}_{(E/B)^2S^2}$
and their $C^{(0)}_{(E/B)^2S^2}$
is $\tfrac12C^{(0)}_{(E/B)^2S^2}$.
The ancillary files use $f_X=2c^S_{10}$, with
$c^S_{10}$ given in Eq.~\eqref{eq:r2s2} of the End Matter
(couplings \texttt{CEE}, \texttt{CEES}, \texttt{CBB},
\texttt{CBBS} for $X=E2$, $ES2$, $B2$, $BS2$).

We obtain the four functions in two independent ways:
from the linearized metric of the companion, through
the eikonal phase, and from its Weyl tensor assembled
from the Newtonian tidal tensor in its rest frame and
contracted with the boosted velocity, integrating the
force directly; the two agree identically.
The same calculation for nonspinning bodies reproduces
the leading tidal angle with electric and magnetic
Love numbers exactly, as given in Eqs.~(4) and (13)
of Ref.~\cite{Cheung:2020sdj} and in the ancillary
file of Ref.~\cite{Jakobsen:2023pvx}.
At leading order in velocity the electric couplings
were checked against Eqs.~(6.19) and (6.20) of
Ref.~\cite{Mandal:2022ufb} and Eq.~(5.23) of
Ref.~\cite{Kim:2022bwv}.
For a test body with these couplings and the curvature
scale held fixed, all four functions were checked
at all velocities against Eq.~(5.19) of
Ref.~\cite{Hoogeveen:2025tew} after mapping its
couplings to ours. A separate calculation with
the couplings on body 2 reproduces the exchange
$\delta\to-\delta$.
At $\sigma=5/4$, $\delta=2/5$ the coefficients of
$C_{X,1}\ah_1^2$ and $C_{X,2}\ah_2^2$ in
$\theta^{(4)}$ are $6.44625464417340$ and
$15.0412608364046$ ($E^2$),
$0.106751859668304$ and $0.249087672559376$
($E^2S^2$),
$4.36573624353246$ and $10.1867179015757$
($B^2$),
$-0.727622707255409$ and $-1.69778631692929$
($B^2S^2$).
The functions are provided in
\texttt{anc/curvature\_squared.py} and
\texttt{anc/CurvatureSquared.m}.

\section{Radiated angular momentum}

With $J^{(3)}_{\rm rad}$ written as in
Eq.~\eqref{eq:jrad} of the Letter and
$X\supset M\sum_S
\big(X^{(0)}_S+\delta X^{(1)}_S\big)S$,
define
$\kappa(s)=\operatorname{arccosh}s/(s^2-1)^{5/2}$
and $\lambda(s)=\log[(s+1)/2]$, with $s=\sigma$.
The polynomials
\begin{align*}
U_5&=119s^5-620s^4-238s^3+648s^2+87s-92,\\
U_6&=119s^6-361s^5-718s^4-746s^3
     +2147s^2-669s+36,\\
U_8&=315s^8+7280s^7-63600s^6+13180s^5
     +183342s^4-274184s^3+311688s^2
     -218212s+45951,\\
V_3&=3s^3-20s^2-3s+4,\\
V_5&=21s^5-78s^4-240s^3-334s^2
     +139s-4,\\
V_7&=315s^7-805s^6+14075s^5+13575s^4
     -33483s^3+11517s^2-35115s+16481,
\end{align*}
give the self-spin functions, for $S=\ah_1^2$,
\begin{align}
X^{(0)}_{\ah_1^2}
&=\frac{3s(2s^2-3)U_5}{32}\,\kappa
-\frac{3U_6}{16(s-1)(s+1)^2}\,\lambda
-\frac{U_8}{640(s^2-1)^2},\\
X^{(1)}_{\ah_1^2}
&=-\frac{21s(2s^2-3)(s^2-1)V_3}{32}\,\kappa
+\frac{3V_5}{16(s+1)^2}\,\lambda
-\frac{V_7}{640(s-1)(s+1)^2},
\end{align}
with $X^{(0)}_{\ah_2^2}=X^{(0)}_{\ah_1^2}$
and $X^{(1)}_{\ah_2^2}=-X^{(1)}_{\ah_1^2}$.

With
\begin{align*}
U^E_5&=77s^5-430s^4-154s^3+444s^2
       +45s-78,\\
U^E_6&=77s^6-293s^5-414s^4-398s^3
       +1149s^2-253s-60,\\
U^E_8&=315s^8-6930s^7+26390s^6+3980s^5
       -56008s^4+19566s^3+2098s^2
       +12568s-4859,\\
V^E_3&=7s^3+70s^2-7s-38,\\
V^E_5&=7s^5+114s^4+90s^3+60s^2
       -129s-30,\\
V^E_7&=315s^7+735s^6-6065s^5-2675s^4
       +3527s^3+21967s^2-16305s+421,
\end{align*}
the quadrupole functions, for
$S=C_{E,1}\ah_1^2$, are
\begin{align}
X^{(0)}_{C_{E,1}\ah_1^2}
&=-\frac{3s(2s^2-3)U^E_5}{32}\,\kappa
+\frac{3U^E_6}{16(s-1)(s+1)^2}\,\lambda
-\frac{U^E_8}{320(s^2-1)^2},\\
X^{(1)}_{C_{E,1}\ah_1^2}
&=-\frac{3s(2s^2-3)(s^2-1)V^E_3}{32}\,\kappa
+\frac{3V^E_5}{16(s+1)^2}\,\lambda
-\frac{V^E_7}{320(s-1)(s+1)^2},
\end{align}
with
$X^{(0)}_{C_{E,2}\ah_2^2}
=X^{(0)}_{C_{E,1}\ah_1^2}$
and
$X^{(1)}_{C_{E,2}\ah_2^2}
=-X^{(1)}_{C_{E,1}\ah_1^2}$.

The mixed-spin function, with
$X^{(1)}_{\ah_1\ah_2}=0$, and the exact static limits
are given in Eq.~\eqref{eq:x2} and End Matter C of the Letter.
Across 80 finite-velocity comparisons, the largest
relative difference between the direct third-PM
calculation and linear-response extraction is
$3.6\times10^{-40}$.

\section{Infrared normalization of the tail logarithm}

The potential--tail cancellation and its rational
and elliptic mechanisms are given in
Eq.~\eqref{eq:poles} and End Matter B.
The normalization of the surviving logarithm
follows from the two-radiative-region factor
$(1-x)^{-4\epsilon}$:
\begin{equation}
\frac{\mathcal P}{2\epsilon}
\big[1-(1-x)^{-4\epsilon}\big]
=2\mathcal P\log(1-x)+\mathcal O(\epsilon).
\end{equation}
Using
$\log[(\sigma-1)/2]=2\log(1-x)-\log(4x)$,
the other finite region terms complete the physical
logarithm $\mathcal P\log[(\sigma-1)/2]$.
The cancellation is exact in velocity; expanding it
near $x=1$ extracts its post-Newtonian coefficient.

\section{Benchmark values}

Table~\ref{tab:mass_functions} gives the two-body
functions of Eq.~\eqref{eq:mass} of the Letter
for the three independent quadratic-spin structures, and
Table~\ref{tab:reference} the complete coefficients of the five
quadratic-spin structures in $\mathcal S_0$ at an asymmetric mass ratio.

\begin{table}[H]
\caption{The two-body functions of
Eq.~\eqref{eq:mass} of the Letter at
$\sigma=5/4$ ($x=1/2$, $v=3/5$),
rounded to 12 significant digits.
The $c^S_{1j}$ are conservative,
the $r^S_{0j}$ are linear in the losses,
and the $r^S_{1j}$ contain parts linear
and quadratic in the losses.
The structures $\ah_2^2$ and
$\CE{2}\ah_2^2$ follow by exchange.}
\label{tab:mass_functions}
\begin{ruledtabular}
\begin{tabular}{lrrrrrr}
$S$ & $c^S_{10}$ & $c^S_{11}$
    & $r^S_{00}$ & $r^S_{01}$
    & $r^S_{10}$ & $r^S_{11}$ \\
\hline
$\ah_1^2$
& $-10665.9129407$
& $925.902632998$
& $3027.11944441$
& $-277.506236056$
& $1337.45415009$
& $-141.546423014$ \\

$\ah_1\ah_2$
& $-13779.7796112$
& $0$
& $6091.82830846$
& $0$
& $2700.09452461$
& $0$ \\

$\CE{1}\ah_1^2$
& $4344.08917066$
& $-203.275862848$
& $-1718.45382208$
& $100.753826566$
& $-763.132544761$
& $48.0578102350$ \\
\end{tabular}
\end{ruledtabular}
\end{table}

\begin{table}[H]
\caption{Coefficients of the displayed monomial in $\theta^{(4)}$ at $x=1/2$ ($\sigma=5/4$) and $\delta=2/5$, rounded to 15 significant digits. The asymmetric mass ratio detects an accidental exchange of the bodies.}
\label{tab:reference}
\begin{ruledtabular}
\begin{tabular}{lrrrr}
$S$ & conservative & linear rad.\ reaction & quadratic rad.\ reaction & total\\
\hline
$\ah_1^2$ & 15443.7381046503 & 634.081985647741 & 2.21548155618658 & 16080.0355718542\\
$\ah_1\ah_2$ & 37344.8591091838 & 1326.69882356786 & 3.56342835104479 & 38675.1213611027\\
$\ah_2^2$ & 21413.4802535141 & 683.168263959465 & 2.23052192078723 & 22098.8790393944\\
$C_{E,1}\ah_1^2$ & $-7397.62582893558$ & $-364.920233706350$ & $-1.53916867242315$ & $-7764.08523131435$\\
$C_{E,2}\ah_2^2$ & $-8854.33831031664$ & $-382.623309897446$ & $-1.55137317681853$ & $-9238.51299339090$\\
\end{tabular}
\end{ruledtabular}
\end{table}

\section{Checks and their coverage}

\textit{Lower spin orders:}
The numerical comparisons at four velocities and
several mass ratios reproduce the spinless conservative
angle to $2\times10^{-37}$, the spinless dissipative terms
to $10^{-40}$ and $2\times10^{-38}$, the conservative
spin-orbit angle to $5\times10^{-37}$, and the
dissipative spin-orbit angle to $3\times10^{-38}$.
The exact expansions reproduce the spinless and
conservative spin-orbit results through $w^{60}$ and
the dissipative spin-orbit result through $w^{30}$.
The closed forms reproduce the published lower-spin
expressions in 96 exact rational-coefficient identities.
At eight velocities (the four production velocities
and $x=33/64,15/29,13/20,36/55$) and five mass
ratios, 960 comparisons with the unrounded direct
evaluation give largest
relative differences of $2.1\times10^{-36}$
(conservative) and $4.6\times10^{-40}$ (dissipative).
Every master-integral value, region and sign that
enters the quadratic-spin angle enters these lower
orders with nonzero weight.

\textit{Quadratic spin:}
The probe part agrees with the expressions in
End Matter A to 43 digits.
For the conservative impulse, exchange symmetry
holds mass sector by mass sector: the potential
coefficients agree as exact rational numbers and
the tail coefficients to 44 digits.

\textit{Numerators and reduction:}
In 192 comparisons, the longitudinal impulse agrees
with its on-shell prediction with largest relative
difference $2.3\times10^{-48}$.
The independent 429-row reduction test described
in End Matter A samples 13 potential and radiation
tables, at rational dimension and velocity points,
after mapping the two reductions to a common master
basis. It does not independently cover every
reduction row.

\end{document}